\documentclass[12pt]{article} 
\usepackage{amsmath,amssymb,color}

\DeclareMathOperator\atanh{atanh}
\def\red#1{{\color{red} #1}}

\let\OLDthebibliography\thebibliography
\renewcommand\thebibliography[1]{
  \OLDthebibliography{#1}
  \setlength{\parskip}{3pt}
  \setlength{\itemsep}{3pt plus 0.3ex}  
}

\begin{document}

\def\prg#1{\par\medskip\noindent{\bf #1}}  \def\ra{\rightarrow}
\newcounter{nbr}
\def\note#1{\bitem\vspace{-5pt}\addtocounter{nbr}{1}
            \item{} #1\vspace{-5pt}
            \eitem}
\def\lra{\leftrightarrow}           \def\Ra{\Rightarrow}
\def\nin{\noindent}                 \def\pd{\partial}
\def\dis{\displaystyle}             \def\Lra{{\Leftrightarrow}}
\def\tgr{{GR$_{||}$}}               \def\tmgl{\hbox{TMG$_\Lambda$}}
\def\vsm{\vspace{-9pt}}             \def\vsmb{\vspace{-5pt}}
\def\cs{{\scriptstyle\rm CS}}       \def\ads3{{\rm AdS$_3$}}
\def\Leff{\hbox{$\mit\L_{\hspace{.6pt}\rm eff}\,$}}
\def\bull{\raise.25ex\hbox{\vrule height.8ex width.8ex}}
\def\Lie{{\cal L}\hspace{-.7em}\raise.25ex\hbox{--}\hspace{.2em}}
\def\sS{\hspace{2pt}S\hspace{-0.83em}\diagup}
\def\ric{{Ric}}
\def\nb{ \marginpar{\bf\Large ?} }  \def\hd{{^\star}}
\def\dis{\displaystyle}             \def\mb#1{\hbox{{\boldmath $#1$}}}
\def\ul#1{\underline{#1}}           \def\ub#1{\underbrace{#1}}
\def\phb{\phantom{\Big|}}
\def\chm{\checkmark}                \def\chmr{\red{\chm}}
\def\ir#1{\,{}^{#1}\hspace{-1.2pt}}
\def\irr#1{\,{}^{(#1)}\hspace{-1.2pt}}

\def\hook{\hbox{\vrule height0pt width4pt depth0.3pt
\vrule height7pt width0.3pt depth0.3pt
\vrule height0pt width2pt depth0pt}\hspace{0.8pt}}
\def\inn{\hook}
\def\first{\rm (1ST)}  \def\second{\hspace{-1cm}\rm (2ND)}

\def\G{\Gamma}        \def\S{\Sigma}        \def\L{{\mit\Lambda}}
\def\D{\Delta}        \def\Th{\Theta}
\def\a{\alpha}        \def\b{\beta}         \def\g{\gamma}
\def\d{\delta}        \def\m{\mu}           \def\n{\nu}
\def\th{\theta}       \def\k{\kappa}        \def\l{\lambda}
\def\vphi{\varphi}    \def\ve{\varepsilon}  \def\vth{{\vartheta}}
\def\p{\pi}
\def\r{\rho}          \def\Om{\Omega}       \def\om{\omega}
\def\s{\sigma}        \def\t{\tau}          \def\eps{\epsilon}
\def\nab{\nabla}      \def\btz{{\rm BTZ}}   \def\heps{{\hat\eps}}

\def\bT{\bar{T}}      \def\hH{\widehat{H}}  \def\hE{\widehat{E}}
\def\tG{{\tilde G}}   \def\cF{{\cal F}}    \def\cA{{\cal A}}
\def\cL{{\cal L}}     \def\cM{{\cal M }}   \def\cE{{\cal E}}
\def\cH{{\cal H}}     \def\hcH{\hat{\cH}}  \def\cT{{\cal T}}
\def\hA{\hat{A}}      \def\hB{\hat{B}}     \def\hK{\hat{K}}
\def\cK{{\cal K}}     \def\hcK{\hat{\cK}}  \def\cT{{\cal T}}
\def\cO{{\cal O}}     \def\hcO{\hat{\cal O}} \def\cV{{\cal V}}
\def\tom{{\tilde\omega}}  \def\cE{{\cal E}} \def\bH{\bar{H}}
\def\cR{{\cal R}}    \def\hR{{\hat R}{}}   \def\hL{{\hat\L}}
\def\tb{{\tilde b}}  \def\tA{{\tilde A}}   \def\hom{{\hat\om}}
\def\tL{{\tilde L}}  \def\tR{{\tilde R}}   \def\tcL{{\tilde\cL}}
\def\he{{\hat e}}    \def\hom{{\hat\om}}   \def\hth{\hat\theta}
\def\hxi{\hat\xi}    \def\hg{\hat g}       \def\hb{{\hat b}}
\def\tH{{\tilde H}}  \def\tV{{\tilde V}}   \def\ha{{\bar a}}
\def\hb{{\bar b}}    \def\bR{\bar{R}}      \def\bF{{\bar F}}
\def\haa{{\bar\a}}   \def\hbb{{\bar\b}}    \def\hgg{{\bar\g}}
\def\tPhi{{\tilde\Phi}} \def\barb{{\bar b}} \def\tPsi{{\tilde\Psi}}
\def\bK{{\bar K}}    \def\bk{{\bar k}}     \def\orth{{\perp}}
\def\bi{{\bar\imath}} \def\bj{{\bar\jmath}} \def \bk{{\bar k}}
\def\bm{{\bar m}}     \def\bn{{\bar n}}    \def\bl{{\bar l}}
\let\Pi\varPi         \def\chH{\check{H}}  \def\bB{{\bar B}}
\let\eR\varOmega    \def\hpi{{\hat\pi}}    \def\hPi{{\hat\Pi}}
\def\cN{{\cal N}}    \def\cB{\cal B}       \def\heps{{\hat\epsilon}}
\def\hN{{\hat N}}    \def\bb{{\bar b}}     \def\fL{L^\text{1st}}
\def\T{\mathbb{T}}

\vfuzz=2pt 
\def\nn{\nonumber}
\def\be{\begin{equation}}             \def\ee{\end{equation}}
\def\ba#1{\begin{array}{#1}}          \def\ea{\end{array}}
\def\bea{\begin{eqnarray} }           \def\eea{\end{eqnarray} }
\def\beann{\begin{eqnarray*} }        \def\eeann{\end{eqnarray*} }
\def\beal{\begin{eqalign}}            \def\eeal{\end{eqalign}}
\def\lab#1{\label{eq:#1}}             \def\eq#1{(\ref{eq:#1})}
\def\balign{\begin{align}}            \def\ealign{\end{align}}
\def\bsubeq{\begin{subequations}}     \def\esubeq{\end{subequations}}
\def\bitem{\begin{itemize}\vspace{-1pt} \setlength\itemsep{-4.5pt} }
  \def\eitem{\end{itemize}\vspace{-1pt} }
\renewcommand{\theequation}{\thesection.\arabic{equation}}
\def\aff#1{\vspace{-12pt}{\normalsize #1}}

\title{Entropy of black holes coupled to a scalar field}

\author{M. Blagojevi\'c and B. Cvetkovi\'c\footnote{
        Email addresses: \texttt{mb@ipb.ac.rs, cbranislav@ipb.ac.rs}} \\
\aff{Institute of Physics, University of Belgrade,
                           Pregrevica 118, 11080 Belgrade, Serbia} }
\date{}
\maketitle

\begin{abstract}
The Hamiltonian approach to black hole entropy, recently  proposed in the framework of Poincar\'e gauge theory, is extended by including the scalar matter. The improved approach is used to analyse asymptotic charges and entropy of a typical black hole with scalar hair in two complementary dynamical
settings, in GR and in the teleparallel gravity. In both cases, the results confirm the validity of the first law.
\end{abstract}


\section{Introduction}

In general relativity (GR), the external field of a charged Kerr-AdS black hole is characterized only by its conserved charges: mass, angular momentum, and electric (magnetic) charge. The idea that this property might be valid for all stationary black holes, often referred to as the ``no-hair" conjecture, had a long-term influence on the black hole physics  \cite{bekenstein-1996,heusler-1996}. Through the years, however, a growing interest in the subject led to a family of the so-called ``hairy" solutions, based on \emph{nonlinear} matter sources coupled to asymptotically flat or AdS black holes \cite{bizon-1994,martinez-2002}. Beside changing our view of the no-hair conjecture, these solutions offer a new insight into dynamical aspects of both the asymptotic charges and black hole entropy.

There exists a Hamiltonian approach to black hole entropy in Poincar\'e gauge theory (PG), proposed in Ref. \cite{mb.bc-2019}. Although the primary idea of that approach was to introduce the concept of entropy for black holes with \emph{torsion} \cite{mb.bc-2020a,mb.bc-2020b}, it can be equally successfully applied to Riemannian black holes, considered as \emph{torsionless} solutions of PG \cite{mb.bc-2020c}. The objective of the present paper is to extend this approach to hairy black holes. Our attention is focussed on the Martinez-Troncoso-Zaneli (MTZ) solution of GR  \cite{martinez-2004} as a characteristic representative of the family of Riemannian black holes with scalar hair.

The paper is organized as follows. In Section 2, we present general dynamical aspects of black holes in PG and give a brief description of entropy as the canonical charge on horizon. In Section 3, we use the Hamiltonian approach to compute energy and entropy, and verify the first law of the MTZ black hole in GR, a subcase of PG defined by $T^i=0$ and the linear-in-curvature action. In Section 4, an analogous analysis is given in the framework of teleparallel gravity (TG), which is a complementary subcase of PG with $R^{ij}=0$. Section 5 is devoted to concluding remarks. In Appendix \ref{appA}, we show that the MTZ black hole is not a solution of the generic PG, and in Appendix \ref{appB}, we derive the scalar matter boundary term.

Our conventions are the same as those in Refs. \cite{mb.bc-2020a,mb.bc-2020b}. Latin indices $(i, j,\dots)$ are the local Lorentz indices, greek indices $(\mu,\nu,\dots)$ are the coordinate indices, and both run over $0,1,2,3$; the orthonormal coframe (tetrad) is $\vth^i=\vth^i{}_\m dx^\m$ (1-form), $\vth=\det(\vth^i{}_\m)$, the dual basis (frame) is $e_i= e_j{}^\m\pd_\m$, and $\om^{ij}=\om^{ij}{}_\m dx^\m$ is the metric compatible connection (1-form); the metric components in the local Lorentz basis are $\eta_{ij}=(1,-1,-1,-1)$, and the totally antisymmetric symbol $\ve_{ijmn}$ is normalized by $\ve_{0123}=1$;
the Hodge dual of a form $\a$ is denoted by $\hd\a$, and the wedge product of forms is implicitly understood
\section{Dynamics and boundary terms}\label{sec2}

We start our exposition with a brief account of the Hamiltonian approach to entropy \cite{mb.bc-2019}, which was introduced in the framework of PG \cite{mb.fh-2013,mb-2002,yo-2006} with the aim of studying entropy of black holes with torsion. However, this formalism can be equally well applied to Riemannian spacetimes. To analyze entropy of the MTZ black hole, found originally in GR \cite{martinez-2004}, we use the tetrad-connection formalism, in which basic dynamical variables are the tetrad field $\vth^i$ and the metric compatible spin connection $\om^{ij}$, and the corresponding field strengths are the torsion $T^i:=d\vth^i+\om^i{}_k\vth^k\equiv\nab\vth^i$ and the curvature $R^{ij}:=d\om^{ij}+\om^i{}_k\om^{kj}$ (2-forms).

The Lagrangian $L$ (4-form) describes the scalar matter coupled to the gravitational field,
\be
L=L_G+L_M\,,                                                      \lab{2.1}
\ee
where $L_G=L_G(\vth^i,T^i,R^{ij})$ is a general (parity even) PG Lagrangian, at most quadratic in the field strengths, and $L_M=L_M(\vth^i,\phi,d\phi)$ describes the scalar matter:
\bsubeq\lab{2.2}
\bea
&&L_G:=-\hd(a_0R+2\L)+T^i\sum_{n=1}^3\hd(a_n\irr{n}T_i)
             +\frac{1}{2}R^{ij}\sum_{n=1}^6\hd(b_n\irr{n}R_{ij})\,,\lab{2.2a}\\
&&L_M:=\frac{1}{2}d\phi(\hd d\phi)+\hd V(\phi)\,.                  \lab{2.2b}
\eea
\esubeq
Here, $a_0=1/(16\pi G)$, $(a_n,b_n,\L)$ are the gravitational coupling constants, $\irr{n}T_i$ and $\irr{n}R_{ij}$ are irreducible parts of the field strengths \cite{mb.fh-2013,yo-2006}, and $V=V(\phi)$ is a self-interaction potential. The field equations are derived by varying $L$ with respect to $\phi$, $\vth^i$, and $\om^{ij}$. By introducing the respective covariant momenta (3-forms)
\be
H_\phi:=\frac{\pd L_M}{\pd d\phi}\,\qquad H_i:=\frac{\pd L_G}{\pd T^i}\,,
\qquad H_{ij}:=\frac{\pd L_G}{\pd R^{ij}}\,,
\ee
these equations can be written in a compact form  as
\bsubeq\lab{2.4}
\bea
&&\cE_\phi:= -dH_\phi+\,\pd_\phi\hd V=0\,,                 \lab{2.4a}\\
&&\cE_i:=\nab H_i+E_i+\t_i=0\,,                                  \lab{2.4b}\\
&&\cE_{ij}:=\nab H_{ij}-(\vth_i H_j-\vth_j H_i)=0\,,                   \lab{2.4c}
\eea
\esubeq
Here, $E_i:=\pd L_G/\pd\vth^i$ and $\t_i:=\pd L_M/\pd\vth^i$ are the gravitational and matter energy-momentum currents (3-forms), respectively.

In the Riemannian limit $T_i=0$, which implies $H_i=0$, the last equation \eq{2.4c} takes the form $\nab H_{ij}=0$, which is, \emph{generically}, for $b_n\ne 0$, not satisfied by the MTZ spacetime; for more details, see Appendix \ref{appA}. Hence, the MTZ black hole is a solution of PG only in the two, geometrically complementary sectors:
\bitem
\item[(s1)] in GR, where $T^i=0$ and also $b_n=0$ (no quadratic curvature terms in $L_G$);
\item[(s2)] in the teleparallel gravity, where $R^{ij}=0$.
\eitem

The Hamiltonian approach to black hole thermodynamics is inspired by the ideas developed by Regge and Teitelboim \cite{regge-1974} and Wald \cite{wald-1993},  but see also Nester and collaborators \cite{chen-1999, chen-2015}. The cornerstone of that approach is the understanding of asymptotic charges (energy and angular momentum) and the horizon charge (entropy) as canonical charges on a 2-dimensional boundary $S$ with two components, one at infinity and the other on horizon, $S=S_\infty\cup S_H$. The related boundary integral $\G:=\G_\infty-\G_H$ is defined by the following variational equations:
\bsubeq\lab{2.5}
\bea
\d\G_\infty&=&\oint_{S_\infty}\d B(\xi)\,,\qquad
       \d\G_H=\oint_{S_H} \d B(\xi)\,,                               \\
\d B(\xi)&:=&(\xi\inn\vth^{i})\d H_i+\d\vth^i(\xi\inn H_i)
   +\frac{1}{2}(\xi\inn\om^{ij})\d H_{ij}
   +\frac{1}{2}\d\om^{ij}(\xi\inn\d H_{ij})                          \nn\\
&&-(\d\phi) (\xi\inn H_\phi)\,.                                     \lab{2.5b}
\eea
\esubeq
For static black holes, the Killing vector $\xi$ has the form $\xi=\pd_t$. The boundary term $\d B$ contains not only the gravitational contribution (upper line), which is sufficient for analyzing \emph{vacuum solutions}, but also an external, matter contribution (bottom line), characterizing \emph{non-vacuum solutions}. In the two sectors (s1) and (s2) defined above, the expression $\d B(\xi)$ is restricted by the conditions  $H_i=0$ and $H_{ij}=0$, respectively.

The operation $\d$ obeys the following requirements:
\bitem
\item[(r1)] it acts on the parameters of a solution, but not on the background configuration;
\item[(r2)] the variation over $S_H$ keeps surface gravity constant.
\eitem
When the boundary integrals $\d\G_\infty$ and $\d\G_H$ are $\d$-integrable and finite, which is controlled by the boundary conditions, they define the asymptotic charges and entropy, respectively.

One should note that the complete boundary term $\G$ is related to the canonical gauge generator $G$ by the variational equation $\d G=-\d\G$ + regular terms \cite{mb.bc-2019,regge-1974}. Hence,  the regularity of $G$ is ensured by the condition
\be
\d\G\equiv\d\G_\infty-\d\G_H=0\,,                                  \lab{2.6}
\ee
which is just the first law of black hole thermodynamics.

In what follows, we will use the Hamiltonian formalism to explore the influence of torsion on the MTZ black hole thermodynamics, by comparing the special cases (s1) and (s2).

\section{The MTZ black hole in GR}\label{sec3}
\setcounter{equation}{0}

\subsection{Geometry}\label{sub31}

Riemannian geometry of the MTZ black hole is determined by the metric \cite{martinez-2004}
\bsubeq\lab{3.1}
\bea
&&ds^2=C^2\Big(N^2dt^2-\frac{dr^2}{N^2}-r^2 d\s^2\Big)\, ,          \nn\\
&&C^2:=\frac{r(r+2Gm)}{(r+Gm)^2}\,,\qquad
  N^2:=\frac{r^2}{\ell^2}-\Big(1+\frac{Gm}{r}\Big)^2\,,
\eea
where $d\s^2$ is the metric of a 2-dimensional manifold $\S$ with constant negative curvature, rescaled to $-1$ for convenience. The manifold $\S$ is locally isometric to the hyperbolic manifold $H^2$, the metric of which can be written in the form
\be
d\s^2=d\r^2+\sinh^2\r\, d\vphi^2\,,
\ee
\esubeq
where $\r\in[0,\infty)$ and $\vphi\in[0,2\pi)$.
Since $H^2$ has infinite area, which is a serious obstacle in thermodynamic considerations, $\S$ is chosen to be of the form $\S=H^2/\G$, where $\G$ is a discrete subgroup of  $SO(1,2)$, the isometry group of $H^2$. This choice ensures the area of $\S$ to be finite \cite{balazs-1986}.
The asymptotic form of $N^2$ suggests that the parameter $m$ is a measure of the black hole mass, whereas its zeros determine the event horizon:
\be
r_+=\frac{\ell}{2}\left(1+\sqrt{1+\frac{4Gm}{\ell}}\,\right)\,.
\ee

Adopting the notation $A:=CN$, $B:=N/C$, $D:=Cr$, one can introduce the orthonormal tetrad associated to the metric \eq{3.1} as
\be
\vth^0=Adt\, ,\qquad \vth^1=\frac{dr}{B}\, ,\qquad
\vth^2=Dd\r\, ,\qquad \vth^3=D\sinh\r\, d\vphi\, ,                \lab{3.3}
\ee
where $\vth^2$ and $\vth^3$ are associated to $H^2/\G$. Then, the horizon area takes the form
\bea
&&A_H=\int_H \vth^2\vth^3=D^2(r_+)\s\,, \qquad
               \s:=\int_H d\r \sinh\r\, d\vphi\,,                    \\
&&D^2\big|_{r_+}:=\ell^2\sqrt{1+\frac{4Gm}{\ell^2}}=\ell(2r_+-\ell)\,.\nn
\eea
The  black hole temperature is defined by surface gravity as
\be
T:=\frac{\k}{2\pi}\,,\qquad
\k:=B\pd_r A\big|_{r_+}
   =\frac{1}{\ell}\sqrt{1+\frac{4Gm}{\ell}}=\frac{1}{\ell^2}(2r_+-\ell)\,.
\qquad
\ee
The Riemannian spin connection reads
\be
\om^{01}=-\frac{A'}{A}B\,\vth^0\,,\qquad \om^{1c}=\frac{D'}{D}B\,\vth^c\,,
\qquad \om^{23}=\frac{\cosh\th}{D\sinh\th}\,\vth^3\,,
\ee
where $c=(2,3)$, and the related curvature is defined by $R^{ij}:=d\om^{ij}+\om^i{}_k\om^{kj}$.

\subsection{Dynamics}

In GR, the field equations \eq{2.4} are restricted by the conditions $T_i=0$ and $H_i=0$. Since $H_{ij}=-2a_0\ve_{ijmn}\vth^m\vth^n$, the vanishing of torsion implies $\nab\vth^i=0$ and consequently $\nab H_{ij}=0$, so that the third field equation \eq{2.4c} is identically satisfied. Hence, dynamical evolution of the MTZ black hole is described by the first two equations,
\bsubeq\lab{3.7}
\bea
&&\cE_\phi=-d(\hd d\phi)+\pd_\phi\hd V=0\,,                 \lab{3.7a}\\
&&\cE_i=E_i+\t_i=0\,.                                             \lab{3.7b}
\eea
\esubeq
Adopting the notation $\m:=Gm$, the solution of the scalar field equation \eq{3.7a} reads \cite{martinez-2004}
\be
\phi:=\sqrt{k}\atanh\Big(\frac{\m}{r+\m}\Big)\,,
\qquad V(\phi):=\frac{k}{\ell^2}\sinh^2\Big(\frac{\phi}{\sqrt{k}}\Big)\,.
\ee
At this stage, the value of the normalization constant $k$ is arbitrary.

To examine the gravitational field equation \eq{3.7b}, we need the explicit expressions for the gravitational and matter energy momentum currents:
\bsubeq
\bea
&&E_i=h_i\inn L_G-\frac{1}{2}(h_i\inn R^{mn})H_{mn}\,,              \\
&&\t_i=h_i\inn L_M-(h_i\inn d\phi)H_\phi\,,
\eea
\esubeq
see, for instance, Refs. \cite{yo-2006,fh.yo-2003}. Then, Eqs. \eq{3.7b} are satisfied for
\be
\L+\frac{3a_0}{\ell^2}=0\,,\qquad k=12a_0=\frac{3}{4\pi}\,.
\ee

\subsection{Boundary charges}

When $H_i=0$, the general boundary term \eq{2.5b} is reduced to
\bea
\d B(\xi)=\frac{1}{2}(\xi\inn\om^{ij})\d H_{ij}
   +\frac{1}{2}\d\om^{ij}(\xi\inn\d H_{ij})
                           -(\d\phi) (\xi\inn H_\phi)\,,            \lab{3.11}
\eea
where $\xi=\pd_t$. Energy and entropy of the MTZ black hole are determined by calculating the corresponding boundary integrals $\d\G_\infty$ and $\d\G_H$, respectively.

\prg{Energy.}
The gravitational and scalar field contributions to the boundary term at infinity are
\bsubeq\lab{3.12}
\bea
\d B_G
 &=&\om^{01}{}_t\d H_{01}+\d\om^{12} H_{12t}+\d\om^{13} H_{13t}  \nn\\
 &=&12a_0\left(-\frac{\m r}{\ell^2}+\frac{4\m^2}{\ell^2}\right)\s\d\m
       +4a_0\s\d\m+O_1\,,  \\
\d B_\phi&=&\-(\d\phi)\phi'AB\vth^2\vth^3=k\d\m\Big(\frac{\m r}{\ell^2}
                       -\frac{4\m^2}{\ell^2}\Big)\s+O_1\,,
\eea
\esubeq
were we use the notation $\xi\inn X^i=X^i{}_t$.
Both terms are divergent, but for $4a_0=3/4\pi$, their sum yields the finite expression for energy:
\be
\d\G_\infty:=\d B_G+\d B_\phi=\frac{\d\m}{4\pi}\s\quad\Ra\quad
        \d E=\frac{\s}{4\pi}\d\m\,.                               \lab{3.13}
\ee

\prg{Entropy.}
On horizon, the boundary contribution $\d B_\phi$ is proportional to the tetrad function $B$, hence it vanishes. Thus, the only nontrivial contribution comes from $\d B_G$,
\be
\d\G_H=\int \d B_G(r_+)=\frac{\s}{4\pi}\d\m\, .                   \lab{3.14}
\ee
Using the explicit expressions for $A_H$ and $T$, one can derive the identity
\be
\d\G_H\equiv T\d S\,,\qquad S:=\frac{A_H}{4\pi}\,,
\ee
which identifies $S$ as the black hole entropy.

\prg{The first law.} The results found in \eq{3.13} and \eq{3.14} ensure the validity of the first law:
\be
\d\G_\infty=\d\G_H \quad\Lra\quad \d E=T\d S\,.
\ee

\section{The MTZ black hole in teleparallel gravity}\label{sec4}
\setcounter{equation}{0}

In the framework of PG, teleparallel gravity is naturally defined by the condition of vanishing curvature, $R^{ij}(\om)=0$, which determines the Lorentz connection $\om^{ij}$ as a pure gauge (unphysical) variable \cite{mb.jn-2020}. The simplest description of TG is obtained by choosing the Lorentz gauge condition $\om^{ij}=0$. As a consequence, the tetrad field $\vth^i$ remains the only dynamical variable, and torsion takes the simple form $T^i=d\vth^i$.

Dynamical properties of TG are determined by the quadratic torsion Lagrangian, inherited from \eq{2.2a},
\be
L_T:=T^i\hd(a_1\irr{1}T^i+ a_2\irr{2}T^i+ a_3\irr{3}T^i)\,.
\ee
When physical predictions of the theory are compared with observational data, one can single out a \emph{one-parameter family} of TG Lagrangians, defined by a single parameter $\g$,
\be
(a_1,a_2,a_3)=a_0\times(1,-2,-1/2+\g)\,,
\ee
which is empirically indistinguishable from GR in the weak field approximation \cite{mb.fh-2013,hs-1979}. In particular, one can show that any spherically symmetric vacuum solution of GR is also a solution of the one-parametr TG.
The case of the MTZ spacetime is somewhat different, as it includes the presence of matter. However, when matter is given as a scalar field, its coupling to   gravity is the same as in GR, and consequently, the MTZ black hole is also a solution of the one-parameter TG coupled to the scalar matter.

To examine thermodynamic properties of the MTZ black hole, we use the  tetrad \eq{3.3} to calculate the teleparallel torsion:
\bea
&&T^0=-\frac{BA'}{A}\vth^0\vth^1\,,\qquad T^2:=\frac{BD'}{D}\vth^1\vth^2\,,\nn\\
&&T^3=\frac{1}{D}\big(\coth\th\,\vth^2\vth^3+BD'\,\vth^1\vth^3\big)\,.
\eea
Since the axial torsion part vanishes, $\irr{3}T^i=0$, the covariant momentum does not depend on the parameter $\g$,
\bsubeq
\be
H_i=2a_0\hd\Big(\irr{1}T^i-2\irr{2}T^i\Big)\,.
\ee
In more details,
\bea
&&H_0=\frac{2a_0}{D}\big(\coth\th\,\vth^1\vth^3
                                -2BD'\,\vth^2\vth^3\big)\,,   \nn\\
&&H_1=-\frac{2a_0}{D}\coth\th\,\vth^0\vth^3\,,                \nn\\
&&H_2=2a_0B\frac{(AD)'}{AD}\vth^0\vth^3\,,\nn\\
&&H_3=-2a_0B\frac{(AD)'}{AD}\vth^0\vth^2\,.
\eea
\esubeq

 In TG, where $H_{ij}=0$, the boundary term \eq{2.5b} is reduced to
\be
\d B(\xi)=(\xi\inn\vth^{i})\d H_i+\d\vth^i(\xi\inn H_i)
           -(\d\phi) (\xi\inn H_\phi)\,.
\ee
For the specific form of the tetrad \eq{3.3}, the gravitational part is simplified to
\be
\d B_G(\xi)=\vth^0{}_t\d H_0+\d\vth^2H_{2t}+\d\vth^3H_{3t}\,.    \lab{4.6}
\ee
\prg{Energy.} At infinity, the gravitational part takes the same value as in GR:
\bsubeq
\bea
\d B_G(\xi)&=&4a_0\s\big[ -AD\d(BD')+BA'D\d D\big]                  \nn\\
 &=&12a_0\s\Big(-\frac{r\m}{\ell^2}+\frac{4\m^2}{\ell^2}\Big)\d\m+4a_0\s\d\m+O_1\,.
\eea
The scalar field contribution also remains the same, see Eq. \eq{3.12}:
\be
\d B_\phi=12a_0\s\Big(\frac{r\m}{\ell^2}-\frac{4\m^2}{\ell^2}\Big)\d\m+O_1\,.
\ee
Summing up, divergent terms cancel each other, and the finite part, which  defines energy, coincides with the corresponding GR value,
\be
\d\G_\infty=\int_{S_\infty}(\d B_G+\d B_\phi)=4a_0\s\d\m\quad\Ra\quad E=\frac{\s}{4\p}\m \,.
\ee
\esubeq

\prg{Entropy.} Since $B=0$ on horizon, the gravitational contribution is simplified by noting that only the terms proportional to $BA'=2\pi T$ do not vanish:
\bsubeq
\be
\d\G_H=\int_{S_H}\d B_G(\xi)=2a_0BA'\int_{S_H}\d(\vth^2\vth^3)=4\pi a_0T\d A_H\,.
\ee
Hence, entropy takes the expected form,
\be
\d\G_H=T\d S\, ,\qquad S:=\frac{A_H}{4}\,.
\ee
\esubeq
\prg{The first law.} The identity $\d\G_\infty=\d\G_H$ ensures the validity of the first law.

\section{Concluding remarks}\label{sec5}
\setcounter{equation}{0}

The Hamiltonian approach to black hole entropy was introduced with the aim of  understanding  entropy of exact vacuum solutions of PG, with or without torsion \cite{mb.bc-2019}. An important generalization of that approach was achieved by applying it to the Ker-Newmann AdS black hole with torsion \cite{mb.bc-2020b}, a classical \emph{non-vacuum} solution of PG. In the present paper, the same approach is used to investigate thermodynamic properties of the MTZ black hole \cite{martinez-2004} as a Riemannian hairy spacetime in the framework of PG.

Since the MTZ black hole is not a solution of the generic PG, it was natural to organize our analysis in the two complementary geometric settings: in GR where $T^i=0$, and in TG where $R^{ij}=0$. In both cases, the values of the boundary terms at infinity (energy) and on horizon (entropy) are calculated and shown to satisfy the first law. The scalar matter boundary term \eq{2.5} is constructed in Appendix \ref{appB}. Although each of the boundary contributions from gravity and matter is diverging at infinity, their sum is finite, ensuring thereby a consistent interpretation of energy in the two sectors.

The expressions for the boundary terms obtained in the GR sector are in agreement with those found in Ref. \cite{martinez-2004}, based on using the Euclidean action formalism. On the other hand, the results obtained in TG cannot be compared to the literature since, as far as we know, there is no any other general approach to black hole entropy in the presence of torsion. The complete agreement of the results obtained in GR with those found in TG indicates a deciding influence of the dynamical factor (the same equations of motion) on thermodynamics, whereas geometric differences (Riemannian versus teleparallel) remain irrelevant.

\section*{Acknowledgements}

We would like to thank Yuri Obukhov for drawing our attention to hairy black holes as a test of the Hamiltonian approach to entropy.
This work was partially supported by the Ministry of Education, Science and Technological development of the Republic of Serbia.

\appendix
\section{On Riemannian solutions of PG}\label{appA}
\setcounter{equation}{0}

In this section, we wish to show that the MTZ black hole is \emph{not a solution} of the generic PG, with all the quadratic curvature terms in $L_G$. For Riemannian solutions, the 2nd, 3rd and 5th irreducible components of the curvature identically vanish \cite{yo-2006}, and the gravitational equation \eq{2.4c} takes the form
\bsubeq
\bea
&&\nab H_{ij}=0\,,                                               \lab{A.1a}\\
&&H_{ij}\equiv -2a_0\hd(\vth^m\vth^n)
     +2\hd\Big[b_1R_{ij}+(b_4-b_1)\irr{4}R_{ij}+(b_6-b_1)\irr{6}R_{ij}\Big]\,,
\eea
\esubeq
where we used $\irr{1}R_{ij}=R_{ij}-\irr{4}R_{ij}-\irr{6}R_{ij}$.
The first term in $H_{ij}$, which represents the GR contribution, has a vanishing divergence since  $\nab \vth^i=0$. The divergence of the second term also vanishes, as follows from the Bianchi identity $\nab R^{ij}=0$. Hence, for the generic (arbitrary) values of the parameters $b_n$, Eq. \eq{A.1a} can be satisfied only if
\be
\nab\,\hd\irr{4}R_{ij}=\nab\,\hd\irr{6}R_{ij}=0\,.               \lab{A.2}
\ee
For our purposes, it is sufficient to analyze the term
\be
\irr{6}R_{ij}=\frac{1}{12}R\,(\vth_i\vth_j)\,.
\ee
Since $\nab \hd\irr{6}R_{ij}=0$ is equivalent to $\nab R=0$, it is evident that the MTZ scalar curvature
\be
R=\frac{12}{\ell^2}
  +6\m^2\frac{\ell^2(\m+r)^4+r^4(3\m+r)(5\m+3r)}{\ell^2r^5(2\m+r^3)}
\ee
does not satisfy the condition \eq{A.2}. This completes the proof.
\section{The scalar field boundary term}\label{appB}
\setcounter{equation}{0}

The canonical charges derived from a local symmetry are defined by boundary terms associated to the canonical generators \cite{regge-1974}. There is a well defined procedure for constructing these generators found by Castellani \cite{castellani-1982}, but it requires a lot of PB calculations. In this section, we use a simpler approach, based on calculating the Noether current and then expressing the field velocities in terms of the canonical momenta.

We start by writing the scalar field Lagrangian \eq{2.2b} in the \emph{first order} form,
\be
\tL_M:=(\pd_\m\phi)P^\m+U(\vth^i{}_\m,P^\m)\,,\qquad
                                U:=-\frac{1}{2\vth}P_\m P^\m+V(\phi)\,,
\ee
where $P_\m$ is an independent dynamical variable, which takes the value $P_\m=\vth\pd_\m\phi$ on shell. The \emph{Noether current} associated to local translations can be expressed in terms of the energy-momentum tensor,
\be
\t^\m:=P^\m\d_0\phi+\xi^\m\tL_M=-\t^\m{}_\l\xi^\l\,,\qquad
\t^\m{}_\n:=P^\m\pd_\n\phi-\d^\m_\n\tL_M\,.
\ee
where we used $\d_0\phi=-\xi^\n\pd_\n\phi$. To switch to the canonical formalism, we replace $P^0$ by the canonical momentum  $\pi:=\pd\tL_M/\pd_0\phi=\vth\pd_0\phi$. Then, $\t^0{}_0=P^0\pd_0\phi-\tL_M$ becomes the \emph{canonical Hamiltonian} $H_c$, whereas the relation $H_c=\vth^i{}_0\cH_i$ implies $\cH_i=\t^0{}_i$.

Being the first class constraints, the Hamiltonian components $\cH_i$ define the generator of local translations,
\be
G_M:=-\xi^\m \vth^i{}_\m\cH_i
    =\xi^\m \vth^i{}_\m\t^0{}_i=-(\xi^\a P^0-\xi^0 P^\a)\pd_\a\phi+\xi^0 U\,,
\ee
where the integration over $d^3x$ is implicit.
Since $G_M$ acts on the phase space variables $\vphi_A$ through the PB operation, $\d \vphi_A:=\{\vphi_A,G_M\}$, it has to be a differentiable functional, that is,  its variation should contain only the terms $\d\vphi_A$, but not $\d\pd\vphi_A$. However, an explicit calculation shows that $G_M$ is not well defined,
\be
\d G_M=-\pd_\a\big[(\xi^\a P^0-\xi^0 P^\a)\d\phi\big]+~ Reg\,,
\ee
where $Reg$ denotes regular, $\d\vphi_A$ terms. Using the identities
\bea
&&(\xi^\a P^0-\xi^0 P^\a)
                =-\frac{1}{2}\ve_{\m\n\b\g}P^\m\xi^\n\ve^{0\a\b\g}\,,\nn\\
&&\xi\inn\hd d\phi=\frac{1}{2}\ve_{\m\n\r\s}P^\m\xi^\n dx^\r dx^\s\,,
\eea
the variation $\d G_M$ can be written in a compact form as
\bea
\d G_M&=&\int_\S\frac{1}{2}
         \pd_\a\big(\ve_{\m\n\b\g}P^\m\xi^\n\d\phi\big)\ve^{0\a\b\g}d^3x+Reg\nn\\
      &=&\int_{\pd\S=S}\d\phi\,\xi\inn H_\phi+Reg\,.
\eea
Denoting this boundary integral as $-\d\G_M$ and moving it to the left hand side, the improved generator $\tilde G_M :=G_M+\G_M$ becomes a well defined functional, whose value represents the scalar matter contribution to energy, for $S=S_\infty$, or to entropy, for $S=S_H$.

\end{document}